\documentclass[twocolumn,showpacs,preprintnumbers,amsmath,amssymb]{revtex4}
\usepackage{graphicx}
\usepackage{dcolumn}
\usepackage{bm}
\begin{document}
\preprint{APS/123-QED}
\title{THE SYSTEMS DYNAMICS OF THE STRUCTURED PARTICLES}
\author{V.M. Somsikov}
 \altaffiliation[] {}
 \email{vmsoms@rambler.ru}
\affiliation{%
Laboratory of Physics of the geoheliocosmic relation, Institute of
Ionosphere, Almaty, Kazahstan.
}%

\date{\today}
\begin{abstract}
Dynamics of the structured particles consisting of potentially
interacting material points is considered in the framework of
classical mechanics. Equations of interaction and motion of
structured particles have been derived. The expression for
friction force has been obtained. It has been shown that
irreversibility of dynamics of structured particles is caused by
increase of their internal energy due to the energy of motion. It
has been shown also that the dynamics of the structured particles
is determined by two types of symmetry: the symmetry of the space
and the internal symmetry of the structured particles. Possibility
of theoretical substantiation of the laws of thermodynamics has
been considered.
\end{abstract}

\pacs{05.45; 02.30.H, J}
\keywords{nonequilibrium, classical mechanics, thermodynamics}
\maketitle

\section{\label{sec:level1}Introduction\protect}

All real bodies in nature are the structured particles ($SP$). But
the existing classical mechanics has been developed for material
points($MP$) or hard bodies [1] which does not exist in the nature
and which have no internal structure. Therefore it is desirable to
create the mechanics of $SP$. This mechanics will be more general
than the existing mechanics of unstructured bodies. Indeed, at the
$MP$ motion in non-homogeneity space and their interaction the
energy of $MP$ motion changes only, while for the $SP$ internal
energy varies also.

As usually the change of $SP$ internal energy is described
empirically by the classical mechanics for $MP$. So a question
arises, whether it is possible to find rigorous mathematical
description of $SP$ dynamics within the frames of the Newtonian
mechanics and if possible then how? We found the answer on this
question by studying the motion equation of $SP$ when $SP$ is an
equilibrium system of potentially interacting $MP$.

It turns out that under certain conditions dynamics of such systems
is irreversible [2-4]. These conditions are formulated as follows:

1). The energy of an $SP$ must be presented as a sum of internal
energy and the energy of $SP$ motion as a whole.

2). Each material point in the system must be connected with a
certain $SP$ independent of its motion in space.

3). During all the process the subsystems are considered to be
equilibrium.

The first condition is necessary to introduce internal energy in
the description of system dynamics as a new key parameter
charactering energy variations of $SP$. The second condition
enables not to redefine $SP$ after mixing of $MP$. The last
condition is taken from thermodynamics. It is equivalent to the
condition of weak interactions in the $SP$, which do not violate
$SP$ equilibrium. Moreover, it implies that each $SP$ contains so
many elements that it can be described using the concept of
equilibrium system.

In this paper we consider derivation of the motion equation of
interacting $SP$. With the help of this equation it is shown how the
mechanism of friction can be explained in the frame of laws of the
classical mechanics. It is shown also how based on the hypothesis of
local equilibrium, which enables to represent non-equilibrium
systems as an ensemble of equilibrium subsystems, one can generalize
the obtained results for two interacting $SP$. It is also shown how
Lagrange, Hamilton and Liouville equations for non-equilibrium
systems are derived from the equation of motion of a set of
equilibrium $SP$. We consider how such equations are different from
their canonic prototypes for the system of $MP$. We consider why the
$SP$ dynamics is determined by the two types of symmetries: the
symmetry of space in which the $SP$ motion and internal symmetry of
distributions of elements of $SP$. It is shown how the main equation
of thermodynamics can be derived from the equation of $SP$
interaction and how the concept of entropy arises in classical
mechanics.

\section{The motion equation of equilibrium structural particles }
It was shown in [4] that for obtaining of the $SP$ motion equation
it is necessary to define the energy of each $SP$ as a sum of
internal energy and energy of its motion. Differentiating energy of
system with respect to time and using a condition of its
conservation, the equation for an energy exchange between $SP$ can
be obtained and then with its help the equation of motion of $SP$
can be found. After that the equation of motion for $SP$ can be
obtained in two stages. At the first stage, based on the condition
of energy conservation, we obtain the equation of motion for the
system in the field of external forces. Then we take a system
consisting of $SP$ and obtain their equations of motion when the
external field for one $SP$ is the field of forces of the other
$SP$. Forces acting between the $SP$ can be obtained from $MP$
potential interaction.

Let us show how the equation of motion for a system of $N$ material
points with weights $m=1$ can be obtained [2-4]. Forces acting
between pairs of $MP$ are assumed to be central and potential. The
energy of the system $E$ is equal to the sum of kinetic energies of
$MP$. Thus $T_N=\sum\limits_{i=1}^{N} m{v_i}^2/2$, their potential
energy in the field of external forces, ${U_N}^{env}$, and potential
energy of their interaction
${U_N}(r_{ij})={\sum\limits_{i=1}^{N-1}}{\sum\limits_{j=i+1}^{N}}U_{ij}(r_{ij})
$, where $r_{ij}=r_i-r_j$, $r_i, v_i$ are coordinates and velocities
of the $i$-th $MP$. Thus, $E=E_N+U^{env}=T_N+U_N+U^{env}=const$.

By substituting variables we represent the energy of the system as a
sum of the motion energy of the center of mass ($CM$) and the
internal energy. Differentiating this energy with respect to time,
we will obtain [3]:
\begin{eqnarray}
V_NM_N\dot{V}_N+{\dot E}_N^{ins}=-V_NF^{env}-\Phi^{env}\label{eqn1}
\end{eqnarray}
Here $F^{env}=\sum\limits_{i=1}^{N}F_i^{env}(R_N,\tilde{r}_i)$,
${\dot E}_N^{ins}={\dot T}_N^{ins}(\tilde{v}_i)+{\dot
U}_N^{ins}(\tilde{r}_i)$=
$\sum\limits_{i=1}^{N}\tilde{v}_i(m\dot{\tilde{v}}_i+F(\tilde{r})_i)$,
 $\Phi^{env}=\sum\limits_{i=1}^{N}\tilde{v}_iF_i^{env}(R_N,\tilde{r}_i)$,
$r_i=R_N+\tilde{r}_i$, $M_N=mN$, $v_i=V_N+\tilde{v}_i$,
$F_i^{env}=\partial{U^{env}}/\partial{\tilde{r}_i)}$, $\tilde{r}_i$,
$\tilde{v}_i$ are the coordinates and velocity of $i$-th $MP$ in the
$CM$ system, $R_N,V_N$ are the coordinates and velocity of the $CM$
system.

The equation (1) represents the balance of the energy of the system
of $MP$ in the field of external forces.

The first term in the left-hand side of the equation determines
the change of kinetic energy of the system -
${\dot{T}}_N^{tr}=V_NM_N\dot{V}_N$. The second term determines the
change of internal energy of the system, ${\dot{E}}_N^{ins}$. This
energy dependent on coordinates and velocities of $MP$ relative to
the $CM$ of the system.

The right-hand side corresponds to the work of internal
forces changing the energy of the system. The first term changes
${\dot{T}}_N^{tr}=V_NM_N\dot{V}_N$. The second term determines the
work of forces changing ${\dot{E}}_N^{ins}$.

Let us determine the condition when the work of non-potential
forces is not equal to zero. We must take into account that
$F^{env}=F^{env}(R+\tilde{r}_i)$ where $R$ is the distance from
the source of force to the $CM$ of the system. Let us assume that
$R>>\tilde{r}_i$. In this case the force $F^{env}$ can be expanded
with respect to a small parameter. Leaving in the expansion terms
of zero and first order we can write:
$F_i^{env}=F_i^{env}|_{R}+(\nabla{F_i^{env}})|_{R}\tilde{r}_i$.
Taking into account that $\sum\limits_{i=1}^{N}\tilde{v}_i
=\sum\limits_{i=1}^{N}\tilde{r}_i=0$ and
$\sum\limits_{i=1}^{N}F_{i}^{env}|_{R}=NF_{i}^{env}|_{R}=F_0^{env}$,
we get from (1):
\begin{eqnarray}
V_N(M_N\dot{V}_N)+
\sum\limits_{i=1}^{N}m\tilde{v}_i(\dot{\tilde{v}}_i+F(\tilde{r})_i)\approx\nonumber\\\approx
-V_NF_0^{env}-({\nabla}F^{env}_{i}|_{R})\sum\limits_{i=1}^{N}\tilde{v}_i\tilde{r}_i\label{eqn2}
\end{eqnarray}

In the right-hand side of equation (2) the force $F_0^{env}$ in the
first term depends on $R$. It is a potential force. The second term
depending on coordinates of $MP$ and their velocities relative to
the $CM$ of the system determines changes in the internal energy of
the system. It is proportional to the divergence of the external
force. Therefore, in spite of the condition $R>>\tilde{r}_i$ the
values of $\tilde{v}_i$ may be not small, and the second term cannot
be omitted. Forces corresponding to this term are not potential
forces. So, the change in the internal energy will be not equal to
zero only if the characteristic scale of inhomogeneities of the
external field is commensurable with the system scale.

Thus, inhomogeneity of space leads to the inhomogeneity of time
for the system. It is connected with possibility of increase in
internal energy of system at the expense of energy of its motion
and impossibility of returning of the system's internal energy
into the energy of its motion due to the law of momentum
conservation. But the law of preservation of full energy is
carried out.

Equation (2) confirms assumption of A. Poincare [5] that it is
necessary to take into account structures of interacting bodies at
rather small distances between them.

Dynamics of an individual $MP$ as well as dynamics of a system of
$MP$ can be derived from equation (1). A $MP$ does not have an
internal energy, and forces acting on it are caused by potential
forces of interaction with other $MP$ and the external force.
Therefore the motion of a $MP$ is determined by the work of
potential forces transforming the energy of the external field into
its kinetic energy only.

Unlike $MP$, a system has its internal energy. Therefore the work of
external forces over the system causes changes in its $T_N^{tr}$ and
$E_N^{ins}$, i.e. the external force breaks up into two components.
The first component is a potential force. It changes momentum of the
system's $CM$. The second component is non-potential. Its work
changes $E_N^{ins}$. Hence, the motion of the system is determined
by the work of potential and non-potential forces transforming the
external field energy into the energy of $CM$ motion and internal
energy.

Multiplying eq.(1) by $V_N$ and dividing by $V_N^2$ we find the
equation of a system motion [4]:
\begin{eqnarray}
M_N\dot{V}_N= -F^{env}-{\alpha_N}V_N\label{eqn3}
\end{eqnarray}

where $\alpha_N=[{\dot E}_N^{ins}+\Phi^{env}]/V_N^2$  is a
coefficient determined by the change of internal energy.

The equation (3) is a motion equation for $SP$. The first term in
the right-hand side of the equation determines the system
acceleration, and the second term determines the change of its
internal energy. The eq. (3) is reduced to the Newton equation if
it is possible to neglect variation in the internal energy.

Thus, the system state in the external field is determined by two
parameters: the energy of motion and the internal energy. Each type
of energy has its own force. The change in the motion energy is
caused by the potential component of the force, whereas the change
in the internal energy is caused by the non-potential component.

Let us show how to obtain the equation for interaction two
equilibrium $SP$. For this purpose we take the system consisting
of two $ES$-$L$ and $K$. The $L$ is the number of elements in the
$L$-$SP$ and $K$ is the number of elements in $K$-$SP$, i.e.
$L+K=N$. Let $LV_L+KV_K=0$, where $V_L$ and $V_K$ are velocities
of $L$ and $K$ equilibrium subsystems relative to the $CM$ of the
system. Differentiating the energy of the system with respect to
time, we obtain:
${\sum\limits_{i=1}^{N}v_i{\dot{v}}_i}+{\sum\limits_{i=1}^{N-1}}\sum\limits_{j=i+1}^{N}v_{ij}
F_{ij}=0$, where $F_{ij}=U_{ij}=\partial{U}/\partial{r_{ij}}$.

In order to derive the equation for $L$-$SP$, in the left-hand
side of the equation we leave only terms determining change of
kinetic and potential energy of interaction of $L$-$SP$ elements
among themselves. All other terms we displace into the right-hand
side of the equation and combine the groups of terms in such a way
that each group contains the terms with identical velocities. In
accordance with Newton equation, the groups which contain terms
with velocities of the elements from $K$-$SP$ are equal to zero.
As a result the right-hand side of the equation will contain only
the terms which determine the interaction of the elements $L$-$SP$
with the elements $K$-$SP$. Thus we will have:
${\sum\limits_{i_L=1}^{L}}v_{i_L}
{\dot{v}}_{i_L}+{\sum\limits_{i_L=1}^{L-1}}\sum\limits_{j_L=i_L+1}^{L}
F_{{i_L}{j_L}}v_{{i_L}{j_L}}={\sum\limits_{i_L=1}^{L}}\sum\limits_{j_K=1}^{K}
F_{{i_L}{j_K}}v_{j_K}$ where double indexes are introduced to
denote that a particle belongs to the corresponding system. If we
make substitution $v_{i_L}=\tilde{v}_{i_L}+V_L$, where
$\tilde{v}_{i_L}$ is the velocity of $i_L$ particle relative to
the $CM$ of $L$ -$SP$, we obtain the equation for $L$-$SP$. The
equation for $K$-$SP$ can be obtained in the same way. The
equations for two interacting systems can be written as [4]:
\begin{eqnarray}
V_LM_L\dot{V}_L+{\dot{E}_L}^{ins}=-{\Phi}_L-V_L{\Psi}
\end{eqnarray}
\begin{eqnarray}
V_KM_K\dot{V}_K+{\dot{E}_K}^{ins}={\Phi}_K+V_K{\Psi}
\end{eqnarray}

Here $M_L=mL, M_K=mK, \Psi=\sum\limits_{{i_L}=1}^LF^K_{i_L}$;
${\Phi}_L=\sum\limits_{{i_L}=1}^L\tilde{v}_{i_L}F^K_{i_L}$;
${\Phi}_K=\sum\limits_{{i_K}=1}^K\tilde{v}_{i_K}F^L_{i_K}$;
$F^K_{i_L}=\sum\limits_{{j_K}=1}^KF_{i_Lj_K}$;
$F^L_{j_K}=\sum\limits_{{i_L}=1}^LF_{i_Lj_K}$;
${\dot{E}_L}^{ins}={\sum\limits_{i_L=1}^{L-1}}\sum\limits_{j_L=i_L+1}^{L}v_{i_Lj_L}
[\frac{{m\dot{v}}_{i_Lj_L}}{L}+\nonumber\\+F_{i_Lj_L}]$;
${\dot{E}_K}^{ins}={\sum\limits_{i_K=1}^{K-1}}\sum\limits_{j_K=i_K+1}^{K}v_{i_Kj_K}
[\frac{{m\dot{v}}_{i_Kj_K}}{K}+\nonumber\\+F_{i_Kj_K}]$.

The equations (4, 5) are equations for interactions two $SP$. They
describe energy exchange between $SP$. Independent variables are
macro-parameters and micro-parameters. Macro-parameters are
coordinates and velocities of the motion of $CM$ of $SP$.
Micro-parameters are relative coordinates and velocities of $MP$.

Therefore the equation of $SP$ interaction binds together two
types of description: on the macrolevel and on the microlevel. The
description on the macrolevel determines dynamics of an $SP$ as a
whole and description on the microlevel determines dynamics of the
elements of an $SP$.

The potential force, $\Psi$, determines the motion of an $SP$ as a
whole. This force is the sum of potential forces acting on the
elements of one $SP$ from the other $SP$.

The forces determined by terms ${\Phi}_L$ and ${\Phi}_K$ transform
the motion energy of $SP$ into their internal energy as a result
of chaotic motion of elements of one $SP$ in the field of forces
of the other $SP$. As in the case of the system in the external
field, these terms are not zero only if the characteristic scale
of inhomogeneity of forces of one system is commeasurable with the
scale of the other system. The work of such forces causes
violation of time symmetry for $SP$ dynamics.

The equations for $SP$ motion corresponding to the equations (4,5)
can be written as [4]:
\begin{equation}
M_L\dot{V}_L=-\Psi-{\alpha}_LV_L \label{eqn6}
\end{equation}
\begin{equation}
M_K\dot{V}_K=\Psi+{\alpha}_KV_K\label{eqn7}
\end{equation}
where ${\alpha}_{L}=(\dot{E}^{ins}_{L}+{\Phi}_{L})/V^2_{L}$,
${\alpha}_{K}=({\Phi}_{K}-\dot{E}^{ins}_{K})/V^2_{K}$,

The equations (6, 7) are motion equations for interacting $SP$.
The second terms in the right-hand side of the equations determine
the forces changing the internal energy of the $SP$. These forces
are equivalent to the friction forces. Their work is a sum of
works of forces acting on the $MP$ of one $SP$ from the other
$SP$.

The coefficients "$\alpha_L$", "$\alpha_K$" determine efficiency
of transformation of the energy of $SP$ motion into their internal
energy. These coefficients are friction coefficients. Therefore
equations (6, 7) enable to determine analytical form of
non-potential forces in the non-equilibrium system causing changes
in the internal energy of the $SP$.

\section{The generals of Lagrange, Hamilton and Liouville equations for equilibrium systems}
Let us show qualitative difference of Lagrange, Hamilton and
Liouville equations for the systems of $MP$ from similar equations
for $SP$.

Using Newton equation one can derive Hamilton principle for $MP$
from differential D'Alambert principle [6]. For this purpose the
time integral of virtual work done by effective forces is equated to
zero. Integration over time is carried out provided that external
forces possess a power function. It means that the canonical
principle of Hamilton is valid only for cases when $\sum F_i\delta
R_i=-\delta U$, where $i$ is a particle number, and $F_i$ is a force
acting on this particle. But for interacting $SP$ the condition of
conservation of forces is not fulfilled because of the presence of a
non-potential component. Therefore Hamiltonian principle for $SP$ as
well as Lagrange, Hamilton and Liouville equations must be derived
using eq. (3).

Liouville equation for non-equilibrium system consisting from a set
of equilibrium $SP$ is written as [2, 4]:
\begin{equation}
df/dt=-\sum\limits_{L=1}^{R}{\partial}{F_L}/{\partial}V_L
\label{eqn8}
\end{equation}

Here $f$ is a distribution function for a set of $SP$, $F_L$ is a
non-potential part of collective forces acting on the $SP$, $V_L$
is the velocity of $L$-$SP$.

The right-hand side of the equation is determined by the
efficiency of transformation of the $SP$ motion energy into their
internal energy. For non-equilibrium systems the right-hand side
is not equal to zero because of non-potentiality of forces
changing the internal energy.

The state of the system as a set of $SP$ can be defined in the phase
space which consists of $6R-1$ coordinates and momentums of $SP$,
where $R$ is the number of $SP$. Location of each $SP$ is given by
three coordinates and their moments. Let us call this space an
$S$-space for $SP$ in order to distinguish it from the usual phase
space for $MP$. Unlike the usual phase space [7,8] the $S$-space is
not conserved. It is caused by transformation of the energy of $SP$
relative motion into their internal energy. The $SP$ internal energy
cannot be transformed into the $SP$ energy of motion as $SP$
momentum cannot change due to the motion of its $MP$ [7]. Therefore
$S$-space is compressible.

\section{The dynamics geometry of $SP$ }
The task of mechanics is definition of trajectories of material
bodies in space with the help of dynamics laws. Therefore the
geometry is included naturally into a formalism of classical
mechanics. The interrelation of geometry and mechanics is carried
out through concept of an interval. This concept lies in bases of
the formalism, both classical, and the relativistic mechanics [6].
We will consider in what difference of an interval for $MP$ from
an interval for $SP$.

Let's consider a point in the configuration space, corresponding to
the system of $MP$. Through an interval time of $dt\longrightarrow0$
the $MP$ will move on distance $ds$. The volume of $ds$ is an
interval. The interval for a set of $MP$ is possible to express
through the kinetic energies as follows [6]:

 \begin{eqnarray}
d\overline{s}^{2} =2T_{N}dt^{2}=\sum^{N}_{i=1}\breve{v}^{2}_{i}dt^{2}=
\sum^{N}_{i=1}(d\breve{x}^{2}_{i}+d\breve{y}^{2}_{i}+d\breve{z}^{2}_{i}) \label{eqn9}
\end{eqnarray}

where ${d\overline{s}}$ is interval displaying infinitesimal
distance between two points of configuration space;
${\breve{x}=\surd{m_ix_i}}$, ${\breve{y}=\surd{m_iy_i}}$,
${\breve{z}=\surd{m_iz_i}}$ are coordinates of the $i$ element;
$m_i$ is a mass of the $i$ -element. The configuration space is
$3N$ dimensional Euclidian spaces for $N$ $MP$. In general case
the linear element will be set in the square-law differential form
of corresponding variables:

\begin{eqnarray}
d\overline{s}^{2}=\sum^{n}_{i,k=1}g_{ik}d\breve{x}_id\breve{x}_k
\label{eqn10}
\end{eqnarray}

where $g_{ik}=g_{ki}$  is symmetrical metrics tensor, $n=3N$.

If we have $p$ kinematics restrictions $f_i=f_i (x_1,x_2...x_n)$,
$i=1,2...p$, the motion of the system will be in $l=3N-p$
dimensional hyperspace. In this case we have:
$d\overline{s}^{2}=\sum^{n}_{i,k=1}a_{ik}dq_idq_k$, where $a_{ik}$
-is known function in a new coordinates. If as kinematics
conditions are potential forces then the equation (8) will be
equivalent to the motion equation of $MP$. But for system which is
a set of $SP$, the energy part is distributed by non-potential
forces. There is a question what will be an interval in this case?

Let's show, that for answer on this question it is necessary to
present energy of system in the form of two parts: energy of
motion of the center of mass of $SP$-$T_N^{tr}$, and internal
energy of $SP$-$T_N^{ins}$. I.e. the interval corresponding for
system $SP$ also should consist of two parts. In this case the
$T_N^{tr}$, $T_N^{ins}$ expressions (7) can be written down as:

\begin{eqnarray}
d\overline{s}^{2}=(2T_N^{tr}+2T_N^{ins})dt^{2}=ds_{tr}^{2}+ds_{ins}^{2}=\nonumber\\\
N\breve{V}_0^2dt^2+(\sum^{N-1}_{i=1}\sum^{N}_{j=i+1}\breve{v}_{ij}^2)dt^2/N
\label{eqn11}
\end{eqnarray}

where $\breve{V}_0=(\sum^{N}_{i=1}\breve{v}_i)/N$,
$\breve{v}_{ij}=\breve{v}_{i}-\breve{v}_{j}$.

Let us transform the energy $T_N$ by replacement:
$\breve{v}_{i}=\breve{V}_{0}-\bar{v}_i$, where
$\sum^{N}_{i=1}\breve{v}_{i}=N\breve{V}_{0}$, i.e.
$\sum^{N}_{i=1}\bar{v}_i=0$. Then we will have:

\begin{eqnarray}
T_N=N\breve{V}_{0}^2/2+\breve{V}_{0}\sum^{N}_{i=1}\bar{v}_i+\sum^{N}_{i=1}\bar{v}_i^2/2
\label{eqn12}
\end{eqnarray}

Because $\sum^{N}_{i=1}\bar{v}_i=0$, then we have

\begin{eqnarray}
\sum^{N}_{i=1}\bar{v}_i^2/2=1/(2N)(\sum^{N-1}_{i=1}\sum^{N}_{j=i+1}\breve{v}_{ij}^2)
\label{eqn13}
\end{eqnarray}

As a result we obtain:

\begin{eqnarray}
d\overline{s}^{2}=(2T_N^{tr}+2T_N^{ins})dt^{2}=ds_{tr}^{2}+ds_{ins}^{2}=\nonumber\\\
N\breve{V}_0^2dt^2+\sum^{N}_{i=1}\bar{v}_i^2dt^2 \label{eqn14}
\end{eqnarray}

Thus, the square of an interval of non-equilibrium system breaks
up to the sum of squares of two intervals. The first corresponds
to the motion energy of $SP$ center of mass and the second
corresponds to the internal energy of system. It is follows from
here that the interval of the non-equilibrium system which
consists of a set of $SP$ breaks up to two independent intervals
characterizing dynamics of system: $ds_{tr}^{2}=
N\breve{V}_0^2dt^2$ and
$ds_{ins}^{2}=\sum^{N}_{i=1}\bar{v}_i^2dt^2$. These intervals are
orthogonally and they correspond to adjacent of a triangle for a
full interval of system in configuration space.

The change of the $SP$ center of mass motion energy is caused by
work of potential forces $F^{tr}$. Their work is defined by
expression: $A^{tr}=\int{F^{tr}dR}$, $F^{tr}=\nabla\varphi$, where
$\varphi$ is scalar function, $dR$ is a distance of systems
motion.

The forces $F^{ins}$ which change of the internal energy $SP$ are
non-potential. Their work consists from the work on change of $MP$
motion energy relative to the center of mass, i.e.
$A^{ins}=\sum^{N}_{i=1}\int{F_idr_i}$, where $dr_i$ -moving of $i$
-th element of system relative to the center of mass. And because
$\sum^{N}_{i=1}{F_i}=0$ then $\int{\sum^{N}_{i=1}F_idR}=0$ for any
possible way of moving of system. I.e. the potential component of
the external force $F^{tr}$ acting on $SP$ changes $s_{tr}$ but
does not change $s_{ins}$. The work of non-potential forces,
$F^{ins}$ changes $s_{ins}$ but does not change $s_{tr}$. The
variables defining motion of the center of mass are
macroparameters, and the variables defining change of internal
energy are microparameters.

Thus for the description of dynamics of the non-equilibrium system
it is necessary to present this system as a set of $SP$ and then
it is necessary to represent $SP$'s energy in the form of the sum
of two types of energy: internal energy and energy of $SP$ motion.
In the nature we deal with the real bodies possessing internal
energy. At their interaction the part of energy go to their
heating. This energy transforming is realized by the friction
force. So the $SP$ dynamics is determined by the two types of
symmetries: the symmetry of space in which the $SP$ motion and
internal symmetry of distributions of elements of $SP$. Thus the
necessity of splitting of the energy on two parts has under itself
a real basis.

\section{The equations of interaction of systems and thermodynamics}

Equations (1-8) give relationship between mechanics and
thermodynamics [4, 8]. According to the basic equation of
thermodynamics the work of external forces acting on the system
splits into two parts. The first part corresponds to reversible
work. In our case it corresponds to the change of the motion energy
of the system as a whole. The second part of energy goes on heating.
It corresponds to the internal energy of the system.

Let us take a motionless non-equilibrium system consisting of "$R$"
equilibrium subsystems. Each equilibrium subsystem consists of a
great number of elements $N_L>>1$, where $L=1,2,3...R,
N=\sum\limits_{L=1}^{R}N_L$. Let $dE$ be work done over the system.
In thermodynamics energy $E$ is called internal energy (in our case
it is equal to the sum of all energies of equilibrium subsystems).
It is known from thermodynamics that ${dE=dQ-PdY}$ [8]. Here,
according to generally accepted terminology, $E$ is the energy of
the system; $Q$ is the thermal energy; $P$ is the pressure; $Y$ is
the volume. The equation of interaction between $SP$ is also a
differential of two types of energy. It means that $dE$ in the $SP$
is redistributed in such a way that some part of it changes energy
of relative motion of the $SP$ and the other part changes the
internal energy. Thus, it follows that entropy may be introduced
into classical mechanics if it is considered as a quantity
characterizing increase in the internal energy  of an $SP$ at the
expense of energy of their motion. Then the increase in entropy can
be written as [3, 4]:
\begin{equation}
{{\Delta{S}}={\sum\limits_{L=1}^R{\{{N_L}
\sum\limits_{k=1}^{N_L}\int[{\sum\limits_s{{F^{L}_{ks}}v_k}/{E^{L}}]{dt}}\}}}}\label{eqn15}
\end{equation}

Here ${E^{L}}$ is the kinetic energy of $L$-$SP$; $N_L$ is the
number of elements in $L$-$SP$; $L=1,2,3...R$; ${R}$ is the number
of $SP$; ${s}$ is the number of external elements which interact
with ${k}$ element belonging to the $L$-$SP$; ${F_{ks}^{L}}$ is
the force acting on the $k$-element; $v_k$ is the velocity of the
$k$- element.

Based on the generally accepted definition of entropy we can derive
expression for its production and define necessary conditions for
stationarity of a nonequilibrium system [4].

\section{Conclusion}
The classical mechanics collides with insuperable difficulties in
attempt to describe evolution of non-equilibrium systems. The main
reason is that the process of evolution is irreversible but the
classical mechanics is reversible [9, 10]. The reversibility of
classical mechanics is defined by the nature of the second law of
Newton. According to this law the acceleration of unstructured
bodies is proportional to the force acting on it. Therefore the
region of application of the second law of Newton is restricted by
unstructured bodies. It means that the second law of Newton is
inapplicable for the description of dynamics of the real bodies
possessing a friction. Hence for removal of the mentioned
restrictions of classical mechanics it is necessary to define
friction forces rigorously on the basis of Newton's second law.

The analysis of dynamics of a hard-discs system has led to the
conclusion that in order to solve this problem it is necessary to
find the motion equation of $SP$. It has been done for a case when
$SP$ represents a system of potentially interacting $MP$, moving
in the field of external forces.

During the process of search of a way which could lead to the $SP$
motions equation and then as a result of its analysis, the
following conclusions were found out.

The motion and evolution of the system are defined by two types of
symmetry: the symmetry of space in which it is moving and its
internal symmetry. In accordance with these two types of
symmetries the energy of system also breaks up on to two types:
the motion energy of system and its internal energy. In its turn,
the change of these types of energy is also defined by two types
of forces. Transformation of energy of $SP$ motion is caused by
potential force. Transformation of internal energy $SP$ is caused
by work of non-potential force. The work of the non-potential
force leads to irreversibility of $SP$ dynamics.

The non-equilibrium systems in approach of the local equilibrium
can be presented as a set of the equilibrium subsystems which are
in motion relative to each other. In this case the description of
dynamics of system by means of the $SP$ motion equation can be
carried out.

The state of the system as a set of $SP$ can be defined in the
phase space which consists of 6R-1 coordinates and momentums of
$SP$, where R is the number of $SP$. Location of each $SP$ is
given by three coordinates and their momentums. The phase space
which is determined by coordinates and velocities of $SP$ is
compressible.

The dynamics of the non-equilibrium system composed of a set of
$SP$ is determined by the Liouville equation for equilibrium $SP$.
These systems acquires an equilibrium state when all energy of
$SP$ motion transforms into its internal energy.

The offered expansion of classical mechanics and the deterministic
explanation of irreversibility open a way to the substantiation of
thermodynamics. According to the motion equation for $SP$ the
first law of thermodynamics follows from the fact that the work of
external forces changes both the energy of particle's motion and
their internal energy. The second law of thermodynamics follows
from irreversible transformation of energy of relative motion of
system's particles into their internal energy.

The motion equation for $SP$ also states impossibility of
existence of structureless particles in classical mechanics, which
is equivalent to infinite divisibility of matter.

Thus, the replacement of model of system in the form of set $MP$
on a model in the form of a set of $SP$ leads to essential
expansion of classical mechanics. Such expansion allows, remaining
within the frame of laws of Newton's mechanics, to offer the
deterministic explanation of irreversibility and, thereby, to
enter the concept of entropy and evolution into the classical
mechanics. It is a bright example of that the further development
of physics is impossible without perfection of models on which
basis it has been constructed.

\smallskip

\end{document}